\documentstyle[preprint,epsf,floats,here,aps]{revtex}

\begin{document}

\draft
 
\title{Singularities in cascade models of the Euler equation} 
\author{C. Uhlig and J. Eggers}
\address{Fachbereich Physik\\ 
  Universit\"at -- Gesamthochschule -- Essen\\D--45117 Essen, Germany}
\date{\today}

\maketitle
 
\begin{abstract}
The formation of singularities in the three-dimensional Euler
equation is investigated. This is done by restricting the 
number of Fourier modes to a set which allows only for local 
interactions in wave number space. Starting from an initial 
large-scale energy distribution, the energy rushes towards smaller
scales, forming a universal front independent of initial 
conditions. The front results in a singularity of the vorticity 
in finite time, and has scaling form as function of the time
difference from the singularity. Using a simplified model,
we compute the values of the exponents and the shape of the front
analytically. The results are in good agreement with numerical 
simulations. 
\end{abstract}


%%%%%%%%%%%%%%%%%
% Introduction  %
%%%%%%%%%%%%%%%%%

\section{Introduction}
\label{sec:intro}

The aim of the theory of fully developed turbulence is to understand 
fluid flow at very high Reynolds numbers. Energy which is fed into
the system at some outer scale $L$ is transported to increasingly
smaller scales through a series of instabilities, until this cascade is 
stopped by the smoothing effect of viscosity. It therefore seems 
natural to consider the limiting equation where the viscosity 
is put to zero, and the Reynolds number thus infinite. The resulting Euler 
equation will not be able to describe a stationary state, where 
the influx of energy is balanced by viscous dissipation. Rather, 
the expectation is that as viscosity no longer limits the smallest
excitable scale, the breakdown of structures will continue 
indefinitely and a singularity of the derivatives of the velocity 
field will appear in finite time, as first suggested by Onsager \cite{o49}.

This singularity has attracted considerable attention, in 
particular from a numerical point of view 
\cite{b83,s85,ps90,gs91,ps92,smo93,k93,b92}.
The reason is twofold: First, the singularity is a reflection
of the instability of turbulent structures, and thus should give 
insight into the mechanism for energy transfer in fully 
developed turbulent flow. Second, as the spatial and temporal
scale of the singular flow gets smaller, one expects the solution
to become independent of boundary or initial conditions, and to reflect
the properties of the nonlinear interaction alone \cite{e93}.
The resulting
solution or class of solutions should thus represent a 
``coherent structure'' of the turbulent flow as well, as long as 
viscosity is not yet important on the scale of its spatial 
variation. Such a structure is a likely candidate to represent 
the small scale structure of a turbulent velocity field, which has
become independent of its outer boundary conditions.

But despite considerable numerical efforts, singularities of the Euler
equation have  
remained elusive. There is disagreement about their expected structure, 
and even the very existence of a singularity is a subject of debate. 
Previous papers have been about equally divided between giving 
indications in favor \cite{s85,gs91,ps92,k93} or against 
\cite{b83,ps90,smo93,b92} the existence of a singularity.
In this paper
we propose to attack the problem using cascade models, which have been 
widely used to study fully developed turbulence. The idea of
cascade or shell models is to divide wavenumber space into 
bands, which cover a certain ratio in wavenumber. Between 
different bands only local interactions are permitted, thus 
implementing the physical idea of local transfer originally 
proposed by Kolmogorov \cite{k41}. This results in a tremendous simplification 
of the problem, both conceptually and numerically. We will thus be able 
to confirm the existence of a singularity unambiguously and to study 
its properties in great detail. Moreover, further simplification 
of the model will allow us to find analytical solutions, which 
confirm the existence of a unique singular shape. This is particularly 
useful since it provides a unified description of both Euler and
Navier-Stokes dynamics in terms of a single cascade model. The solid
understanding of both aspects should enable us to ascertain the 
significance of Euler singularities to turbulent flow. 

In the next section we will introduce a class of models originally 
developed in \cite{eg91}, known as {\bf RE}duced {\bf W}ave vector
set {\bf A}pproximations or REWA models. They arise by 
restricting the number of available Fourier modes to a self-similar
set, with a constant number of modes within an octave in wave-number.
The properties of REWA models have been studied extensively in the 
context of stationary turbulent flow \cite{eg91,gl94,ue96}.
In particular, it was shown \cite{ue96} that the turbulent 
fluctuations are characterized by a set of anomalous scaling 
exponents, as suggested by the multifractal theory of turbulence
\cite{nelkin89}. Here we make the connection between the inviscid 
singularity and the stationary state of turbulent flow by 
presenting a simulation of decaying turbulence, starting from 
an initial large-scale distribution of energy. If the viscosity is 
sufficiently small, the flow will be effectively inviscid, resulting
in a rapid build-up of velocity gradients. Eventually, after sufficiently
small scales are excited, viscosity becomes important, leading to 
dissipation of energy. As inertial transport and viscous damping 
balance, the energy spectrum becomes flatter and close to a 
Kolmogorov spectrum.

In the third section we study the formation of singularities for 
very long cascades at zero viscosity. Starting from arbitrary 
initial conditions, a universal front develops, which is self-similar:
at different time distances from the singularity the solutions
can be collapsed by a rescaling of their length scale. The smallest 
excited scale $r_c$ follows a power law as function of the 
distance $\tau = t^{\ast} - t$ from the singularity: 
$r_c \sim \tau^{\beta}$. The relevant exponents and the 
form of the front are determined. In the fourth section we
develop an analytical description of the singularity by using an
effective equation for the energy of the shell. The same effective 
equation has been used before \cite{e94,ue96} to compute the anomalous
exponents of stationary turbulence. We show that fluctuations are 
irrelevant for the description of the Euler singularity. The resulting 
deterministic equation can be reduced to an ordinary differential 
equation if the self-similarity of singular solutions is 
exploited. 

This ordinary differential equation is used in the fifth section 
to study the selection of universal solutions out of arbitrary 
initial data. There exists a family of solutions parameterized 
by the exponent $\beta$, which connects length and time scales. 
For large $\beta$, the solutions develop unstable fronts, which 
are unphysical. Thus the most singular solution which is not yet
unstable is selected. The resulting unique solution agrees 
well with numerical simulations of the REWA cascade. In the 
discussion we comment on related work and point out possible uses 
of the present study of inviscid singularities for the understanding
of fully developed turbulence.

%%%%%%%%%%%%%%%%%
% section 2     %
%%%%%%%%%%%%%%%%%

\section{Model equations}
\label{sec:model}

A variety of shell models have been proposed in the past to 
describe a turbulent cascade 
\cite{obukhov71,gledzer73,yamada87,jensen91,eg91}. By using a 
dynamical model, one hopes to gain insight into the origin 
and the statistics of turbulent fluctuations. Since a cascade 
model consists of a linear structure of turbulence elements, 
the problem is simplified enormously, both from an analytical 
and a computational point of view. However, investigations
of cascade models have been limited to the steady state,
where energy input equals dissipation on the average. The question 
addressed here is whether cascade models are also capable of
describing some of the instabilities of inviscid flow, which 
lead to the build-up of gradients. 

The cascade model we consider here was introduced in 
\cite{eg91}, and is sometimes called the REWA model. It has been 
used extensively to study the stationary state of fully 
developed turbulence \cite{eg91,gl94,ue96}. It is based on the full 
Fourier-transformed Navier-Stokes equation with a volume of 
periodicity $(2\pi L)^3$. Only local interactions are taken 
into account, which is implemented by projecting the Navier-Stokes 
equation onto a self-similar set of wave numbers ${\cal K}=\bigcup_{\ell}{\cal
K}_{\ell}$. Each of the wave vector shells ${\cal K}_{\ell}$ represents
an octave in wave number, which greatly reduces the total
number of modes, making the model numerically tractable. 
The shell ${\cal K}_0$ describes the motion of the largest 
elements in the flow, which are of the order of the outer length
$L$. It is composed of $N$ wave vectors ${\bf k}^{(0)}_i :
{\cal K}_{0}=\left\{{\bf k}^{(0)}_{i}:i=1,\dots,N\right\}$.
Starting with the generating shell ${\cal K}_0$, the other 
shells are found rescaling ${\cal K}_0$ with a factor of 2:
${\cal K}_{\ell} = 2^{\ell} {\cal K}_0$. The shell ${\cal K}_{\ell}$
thus represents structures of size $r \sim 2^{-\ell} L$. In a 
turbulent cascade, this scaling procedure is followed until one
reaches a Kolmogorov length $\eta$, where the turbulent motion
is damped by viscosity. In the present paper, we will mostly be
concerned with the limit of zero viscosity. Thus arbitrarily
small scales can be excited, and our simulations are valid only 
for a finite time, until energy is transferred into the smallest
scale available. By choosing the number of levels very large, we are 
still able to extract reliable scaling information. 

Explicitly, the projection of the Navier-Stokes equation reads
\begin{mathletters}
  \label{FWNS}
  \begin{eqnarray}
    \frac{\partial}{\partial t}u_{i}({\bf k},t)&=&
    -\imath M_{ijk}({\bf k})\sum_{
      {\bf p},{\bf q}\in{\cal K}_{\ell}\atop {\bf k}={\bf p}+{\bf q}} 
    u_{j}({\bf p},t)u_{k}({\bf q},t)-\nu k^{2}u_{i}({\bf k},t)
    \label{FWNSa}\\
    {\bf k}\cdot {\bf u}({\bf k},t)&=&0\label{FWNSb}\quad .
  \end{eqnarray}
\end{mathletters}
The coupling tensor 
$M_{ijk}({\bf  k}) =
\left[k_{j}P_{ik}({\bf k})+k_{k}P_{ij}({\bf k})\right]/2$ with the
projector $P_{ik}({\bf k})=\delta_{ik}-k_{i}k_{k}/k^{2}$ is symmetric 
in $j,k$. The inertial part of (\ref{FWNSa}) consists of all 
triadic interactions modes with ${\bf k = p + q}$. For the 
moment we have kept the viscous term, but will put $\nu = 0$ 
later for our study of the Euler equation. With this approximation 
the energy of a shell is 
\begin{equation}
  \label{energy}
  E_{\ell}(t)=\frac{1}{2}\sum_{{\bf k}\in{\cal K}_{\ell}} |{\bf u}({\bf
    k},t)|^{2}\quad .
\end{equation}
Before the appearance of a singularity the total energy 
\begin{equation}
  \label{Etot}
E_{tot}(t)=\sum_{\ell=0}^{\infty}E_{\ell}(t)
\end{equation}
is conserved for $\nu = 0$. 

In an earlier paper \cite{ue96} we have investigated the effect 
of different choices for the wave vector set ${\cal K}_0$ in
some detail. Here we will mostly deal with a single set of $N = 26$
modes, where the components of ${\bf k}^{(0)}_i$ consist of
all combinations of $0, -1$, and $1$, because we found the properties of the 
inviscid singularity to be quite insensitive to 
the specific choice of ${\cal K}_0$. The ${\cal K}_0$ 
considered here only allows for local couplings between 
shells, and thus the energy is transported only between 
adjacent shells. Hence if $T_{\ell\rightarrow\ell+1}(t)$
is the energy transfer from shell $\ell$ to $\ell+1$, and no dissipation 
occurs, we can write an energy balance equation
\begin{equation}
  \label{enbal}
  \frac{d}{dt}E_{\ell}(t) =
    T_{\ell-1\to\ell}(t) -T_{\ell\to\ell+1}(t) \quad .
\end{equation}
The transfer $T_{\ell\rightarrow\ell+1}(t)$ can be written explicitly
as a sum over triple products of velocity modes. Equation (\ref{enbal})
will be the basis for a simplified description of the cascade, which 
we will use later to obtain analytical solutions. 

\begin{figure}[H]
  \begin{center}
    \leavevmode
    \epsfsize=0.5 \textwidth
    \epsffile{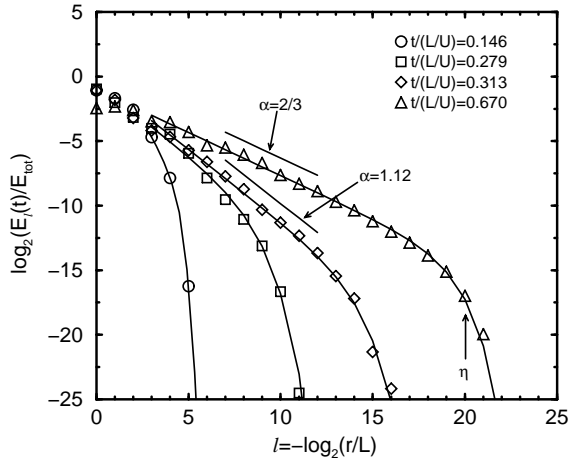}    
  \end{center}
  \caption{Evolution of the REWA cascade at a Reynolds number 
    of $1.35\cdot10^8$. The energy is initially localized in the $\ell=0$
    level. Viscous effects are small at first, and a scaling 
    regime with exponent $\alpha = 1.12$ develops. After the
    singular front is stopped by viscosity, the energy levels
    off to a Kolmogorov spectrum.
      }
  \label{fig:decay}
\end{figure}

We illustrate the formation of a singularity in our model by considering 
the dynamics (\ref{FWNS}) with an initial condition where only 
the modes of level $\ell=0$ are excited. To make a connection with 
stationary turbulence, we keep $\nu$ small but finite in this 
example. The Reynolds number 
\[
Re = \frac{L U}{\nu}, 
\]
where $U$ is the typical amplitude of a velocity mode on the 
highest level, is $1.35\cdot10^8$. Figure \ref{fig:decay}
shows the resulting evolution of the shell energies. 
Energy rushes downward in scale to fill the shells which are
not yet excited. As shown in \cite{ue96}, this can be seen 
as a result of the tendency of the dynamics to establish equipartition
of energy between shells. Since the small scales are not excited at all,
the energy transfer is directed almost exclusively towards smaller 
scales. This causes a front to form, beyond which no excitation 
has yet taken place. As this front penetrates the small scale 
regions, it leaves behind a power law distribution of the energy, 
whose exponent is close to $\alpha = 1.12$. Eventually the front 
feels the viscosity, which happens at the Kolmogorov length $\eta$,
estimated from the initial conditions and the viscosity. 
Since the energy is now dissipated instead of transferred, the 
front stalls, and an equilibrium of inertial transfer and 
energy dissipation is established. Now there is also significant 
backflow of energy, and a transfer towards smaller 
scales is observed only on the average. Thus the profile gradually 
reduces in steepness and converges to the familiar Kolmogorov 
form \cite{k41,f95}, with a scaling exponent close to the classical value of 
$2/3$. Since there is no energy input, a truly stationary state
cannot be established, and all excitations will decay to zero in the 
infinite time limit. This however will happen on much longer time 
scales than seen in Fig. \ref{fig:decay}. 

From this we observe that a Kolmogorov state develops from the interplay 
between singular motion and viscous dissipation.
We will now concentrate on the early time behavior,
where viscosity is not yet important. Our aim is to explain
the value of the scaling exponent $\alpha > 2/3$, and to find the 
structure of the singular front.

%%%%%%%%%%%%%%%%%
% section 3     %
%%%%%%%%%%%%%%%%%

\section{The Euler singularity}
\label{sec:euler}

Here we describe the evolution of very long cascades, where the 
viscosity has been turned off. Since we only look at
a single trajectory leading up to the singularity, and no statistics
have to be accumulated, we can easily 
afford to simulate 100 levels, corresponding to 30 orders
of magnitude in scale. This will allow us to
identify the scaling behavior of the singularity unambiguously. 

\begin{figure}[H]
  \begin{center}
    \leavevmode
    \epsfsize=0.5 \textwidth
    \epsffile{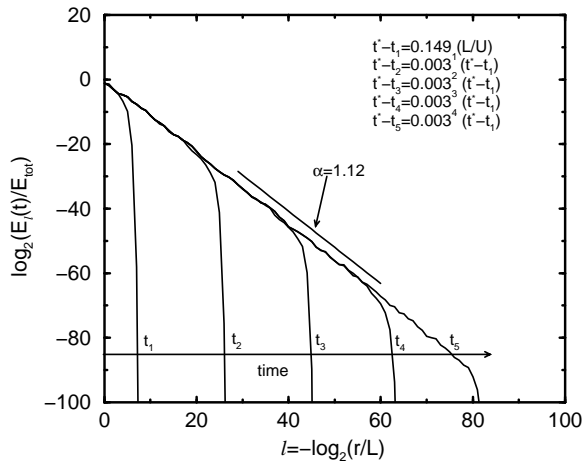}    
    \caption{The energy cascade for the REWA model with 100 levels.
     Energy is concentrated in the top level initially.
     The energy drops to zero at a finite level number. This 
     position moves in time intervals which scale geometrically  
     as the singularity is approached.
                 }
    \label{fig:FWen}
  \end{center}
\end{figure}

The result of a simulation, where again only the level $\ell=0$
is excited, is shown in Fig.\ref{fig:FWen}. Note that the axes
are logarithmic, so the small scales contain only a very small 
fraction of the energy.
\noindent
The distribution of energies $E_{\ell}$ 
looks very similar to the previous figure, except that
the absence of viscosity allows the cascading to continue indefinitely. 
Since the length scale associated with 
level $\ell$ is $r = 2^{-\ell} L$, the energy spectrum behind the 
front is a power law
\begin{equation}
  \label{alpha}
E_{\ell} \sim r^{\alpha} \quad,
\alpha = 1.12 \pm 0.01 \quad.
\end{equation}
At any given time, the energy drops to zero at the front. The scale 
$r_c$ where this happens thus represents the smallest excited 
scale, which goes to zero at a finite time $t^{\ast}$ which 
depends on initial conditions. Thus one expects sufficiently
high derivatives of the velocity field 
to blow up as the time difference from 
the singularity 
\begin{equation}
   \label{tau}
\tau = t^{\ast} - t
\end{equation}
goes to zero. Indeed, plotting the smallest excited scale $r_c$
as a function of $\tau$, one again finds a power law
\begin{figure}[H]
  \begin{center}
    \leavevmode
    \epsfsize=0.5 \textwidth
    \epsffile{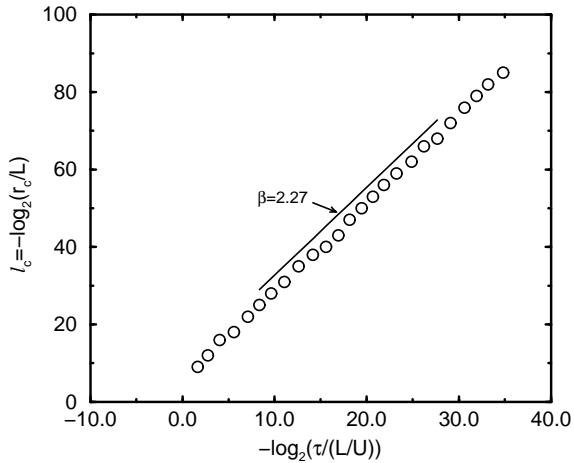}    
    \caption{The level where the energy drops to zero, as 
     function of the time distance from the singularity. 
     The exponent $\beta$ is related to 
     $\alpha$ by the scaling relation $\alpha = 2 - 2/\beta$.
           }
    \label{fig:r-t}
  \end{center}
\end{figure}
\begin{equation}
   \label{beta}
r_c \sim \tau^{\beta} \quad,
\beta = 2.27 \pm 0.01 \quad,
\end{equation}
see Fig. \ref{fig:r-t}
A relation between the two power laws (\ref{alpha}) and (\ref{beta})
is established by the following argument: the time scale on which the 
singularity is moving must be $\tau$ itself, so a typical energy 
at the front is $(r_c/\tau)^2 \sim r_c^{2(\beta-1)/\beta}$. 
Comparing this with the scaling law (\ref{alpha}) one finds the 
scaling relation 
\begin{equation}
  \label{scaling}
\alpha = \frac{2(\beta - 1)}{\beta} \quad,
\end{equation}
which is obeyed precisely by the values found for $\alpha$ and
$\beta$ numerically. 

\begin{figure}[H]
  \begin{center}
    \leavevmode
    \epsfsize=0.5 \textwidth
    \epsffile{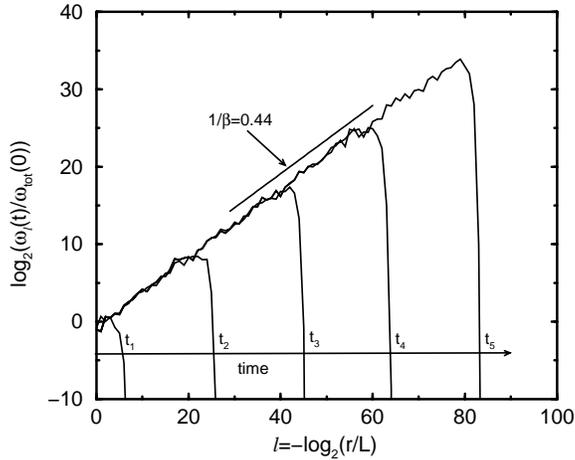}
    \caption{The vorticity $\omega_{\ell}$ within a shell 
    $\ell$ as function of level number. The maximum value,
     which is reached at the front, diverges like $\tau^{-1}$ 
      in time.
          }
    \label{fig:vort}
  \end{center}
\end{figure}

The typical velocity is from (\ref{alpha}) expected to go down
like $r^{\alpha/2}$ in scale. Thus the vorticity of a shell $\ell$,
defined by 
\begin{equation}
  \label{omega}
  \omega_{\ell}(t) = \left|\sum_{{\bf k}\in {\cal K}_{\ell}} {\bf k}\times
  {\bf u}({\bf k},t)\right|\quad ,
\end{equation}
behaves like $r^{\alpha/2-1} = r^{-1/\beta}$ and reaches its maximum
near the front, as seen in Fig. \ref{fig:vort}.
\noindent
Since $\omega$ has units of inverse time, this maximum 
diverges like $\tau^{-1}$, which we confirmed numerically.
This means the singularity observed 
in the REWA model is consistent with the criterion by 
Beale, Kato, and Majda \cite{bkm84}, that the maximum of 
the vorticity should diverge at least as fast as $\tau^{-1}$ 
for a true Euler singularity. Because the exponent of the vorticity
is known, we used the scaling relation
\begin{equation}
  \label{bkm}
\max_{\ell} \{\omega_{\ell}\} \sim (t^{\ast} - t)^{-1}
\end{equation}
to fit the value of the singular time $t^{\ast}$.

\begin{figure}[H]
  \begin{center}
    \leavevmode
    \epsfsize=0.5 \textwidth
    \epsffile{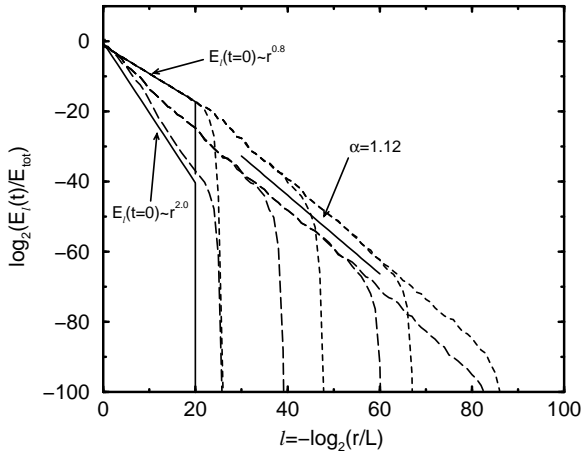}    
    \caption{The singularity for two different initial conditions.
    If the exponent is smaller than 2 initially, only the front
    moves. If it is 2, all scales move on the same time scale initially.
            }
    \label{fig:initial}
  \end{center}
\end{figure}

Next we look at the possible influence of initial conditions on the 
singularity. In Fig. \ref{fig:initial}
\noindent
we chose the energy to have a nontrivial distribution at the initial
time. For this distribution we chose two power laws, one with an
exponent smaller than $\alpha$, the other larger. As seen in the 
figure, in both cases the solution settles on the same slope, 
with the same universal shape at the front. We can thus conclude that 
the singularity is universal except for the value of the singular
time $t^{\ast}$ and the energy scale. This is 
of course only true apart from small fluctuations of the energy.
These result from the complicated chaotic motion 
of Fourier modes underlying the excitation of small scales. 
However at the front one is very far from equilibrium, so fluctuations 
in the energy transfer are small compared with its absolute 
value.

There is another interesting observation to be made in Fig. \ref{fig:initial},
which hints to the observed universality. If the exponent of the 
energy distribution is $0.8$ initially, only the front of the 
distribution moves, the contribution from larger scales remains
static. This is because $\tau_r = (r^2/E)^{1/2} = r^{1-\alpha/2}$ represents a 
local time scale. Hence as long as $\alpha < 2$, only the 
smallest available scale moves, since it has the shortest time scale. 
The limiting case $\alpha = 2$ is the other initial distribution 
given in Fig. \ref{fig:initial}, and indeed it now evolves on all 
length scales. But as soon as the newly formed front overtakes 
the old one, it again only grows from its front, since the slope 
behind it is now smaller than $2$. Hence in each case universality results 
from the fact that growth is determined only from the local properties
of the front. 

\begin{figure}[H]
  \begin{center}
    \leavevmode
    \epsfsize=0.5 \textwidth
    \epsffile{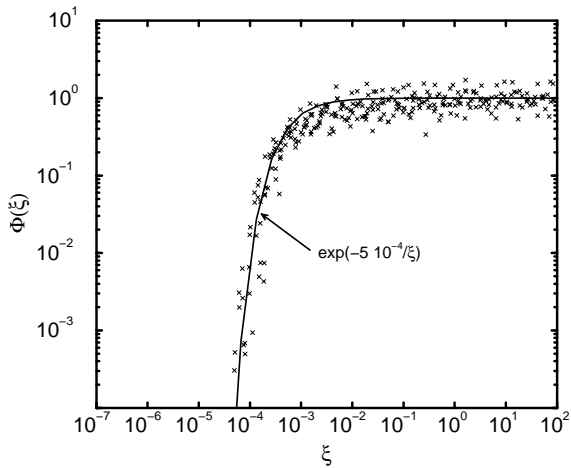}    
    \caption{The scaling function of the energy $\Phi(\xi)$ 
    determined from simulations of the REWA cascade using
    the rescaling (\protect{\ref{scalform}}). It is well 
    fitted by a functional form proposed originally 
    \protect{\cite{b83}} for singularities of the full 
    Euler equation.
            }
    \label{fig:phi}
  \end{center}
\end{figure}

Having seen that the characteristic length and time scales of the 
singularity behave like power laws, we see next whether the 
whole sequence of profiles can be rescaled to fall onto a single 
master curve. To that end, we first nondimensionalize length and
time. If $L$ is a length where the energy already has its scaling 
form, and $E_0$ is the energy on that scale, one can introduce the 
nondimensional quantities 
\begin{equation}
  \label{refscal}
  \tilde r = r/L ,\ \tilde E_{\ell} = E_{\ell}/E_0,\ \tilde t
  = t (E_0/L^2)^{1/2} \quad.
\end{equation}
Using these, the energy is expected to scale like 
\begin{equation}
  \label{scalform}
  \tilde E_{\ell}(\tilde\tau)={\tilde r}^{\alpha}\Phi(\xi)\quad ,
  \xi = \frac{\tilde r}{{\tilde \tau}^{\beta}} \quad,
\end{equation}
where $\Phi$ is a universal function. In Fig. \ref{fig:phi}
\noindent
the energies at different distances from the singularity 
are superimposed according to (\ref{scalform}). The values of the energy 
at different levels is marked by crosses. Allowing for some fluctuations,
the scaling relation is obeyed very well. Owing to the rescaling 
(\ref{refscal}),(\ref{scalform}), 
$\Phi$ asymptotes to $1$ as $\xi\rightarrow\infty$,
and the collapse is the same for all initial conditions. Finally we note
that $\Phi$ is fitted very well by 
\begin{equation}
  \label{frisch}
   \Phi(\xi) = \exp(-5\cdot10^{-4}/\xi) \quad,
\end{equation}
which is a functional form of the energy spectrum proposed by 
Brachet et al. \cite{b83}. We will comment on the relation 
between our work and \cite{b83} in the discussion.

%%%%%%%%%%%%%%%%%
% section 4     %
%%%%%%%%%%%%%%%%%

\section{Towards an analytical description}
\label{sec:sim}

We now develop an analytical theory for the form of the singularity
which will also explain its universality, i.e. independence of 
initial conditions. As a basis we use a simple model for the energy 
transfer \cite{e94}, which has been used before to describe the 
stationary state of developed turbulence \cite{ue96}. The energy 
transfer is split into two parts, one deterministic, the other
stochastic:
\begin{equation}
  \label{langevin}
T_{\ell\rightarrow\ell+1}(t) = T^{(det)}_{\ell\rightarrow\ell+1}(t) + 
              T^{(stoch)}_{\ell\rightarrow\ell+1}(t) \quad.
\end{equation}
Once $T_{\ell\rightarrow\ell+1}(t)$ is specified, conservation of energy 
(\ref{enbal}) results in an equation of motion for the energy. 
The deterministic part expresses the tendency of the cascade to
establish equipartition of energy between its members. The stochastic part 
represents the chaotic mixing of the individual Fourier modes. 
If in addition we assume that both $T^{(det)}_{\ell\rightarrow\ell+1}(t)$ 
and $T^{(stoch)}_{\ell\rightarrow\ell+1}(t)$ only depend on the neighboring 
values of the energies, one ends up with the expressions \cite{ue96}
\begin{mathletters}
  \label{Lt}
  \begin{eqnarray}
    T_{\ell\rightarrow \ell+1}^{(det)}(t)&=& D
    \frac{2^{\ell}}{L}\left(E_{\ell}^{3/2}(t)-E_{\ell+1}^{3/2}(t)\right),
    \label{Lta}\\ 
    T_{\ell\rightarrow \ell+1}^{(stoch)}(t)&=& R
    \left(\frac{2^{(\ell+1)}}{L}\right)^{1/2}
      (E_{\ell}(t)E_{\ell+1}(t))^{5/8}
      \xi_{\ell+1}(t) \quad. \label{Ltb}
  \end{eqnarray}
\end{mathletters}
The powers appearing in (\ref{Lt}) are derived from dimensional
considerations, and $\xi$ represents a Gaussian white noise with
$\left<\xi_{\ell}(t)\right>=0$ and
$\left<\xi_{\ell}(t)\xi_{\ell'}(t')\right>=2\delta_{\ell\ell'}\delta(t-t')$.
We use Ito's definition in equation (\ref{Ltb}). Together with 
(\ref{enbal}), (\ref{Lt}) is a Langevin equation for the motion of
a cascade, so we will refer to it as the Langevin model.
It has been shown that the model given by (\ref{enbal}),(\ref{Lt}) 
exhibits multifractal 
scaling in a stationary turbulent state, and anomalous scaling exponents 
can be calculated analytically \cite{e94}. At the same time it gives 
an excellent description of the turbulent state of the REWA cascade
\cite{ue96}. 

\begin{figure}[H]
  \begin{center}
    \leavevmode
    \epsfsize=0.5 \textwidth
    \epsffile{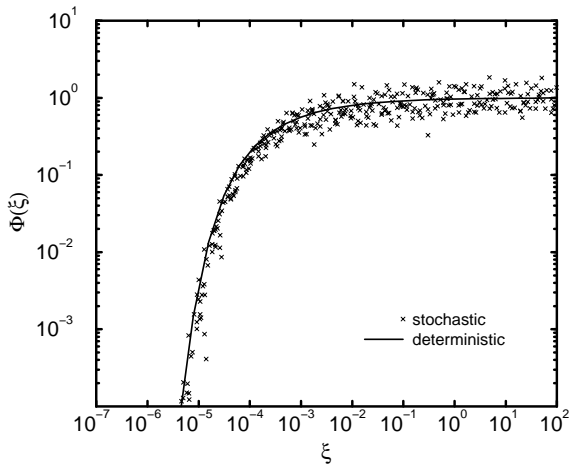}        
    \caption{The same scaling function $\Phi$ as in 
    Figure \protect{\ref{fig:phi}}, but for the Langevin 
    model (\protect{\ref{enbal}}),(\protect{\ref{Lt}}).
    The crosses represent the full model, the full line 
    corresponds to the deterministic case $R = 0$.
       }
    \label{fig:phi-L}
  \end{center}
\end{figure}
Figure \ref{fig:phi-L} shows a simulation of the model equations 
(\ref{enbal}),(\ref{Lt}) at zero viscosity, 
with energy concentrated in the largest 
scale initially. The profiles have been rescaled as in (\ref{scalform}),
which again leads to a collapse very similar to that of Fig. 
\ref{fig:phi}. The exponent $\alpha = 1.24$ is slightly 
larger than that found for the REWA cascade. The values of the 
free parameters $D$ and $R$ in (\ref{Lt}) are taken from \cite{ue96}:
\begin{equation}
  \label{parameters}
 D = 6.5\cdot 10^{-2} \quad, R = 4.4 \cdot 10^{-2} \quad.
\end{equation}
The amplitude $D$ is of particular significance, since it determines 
the effectiveness of the energy transfer. If $D$ gets larger, 
the front reaches a given scale
at an earlier time, as we are going to see in more detail below.

On the other hand, the noise strength $R$ is insignificant for the 
formation of the singularity, as fluctuations are small, in agreement
with the result found for the REWA cascade. This is to 
be expected since the motion of the singularity is dictated 
by the front which is very steep. Thus the 
deterministic part (\ref{Lta}), which consists of an energy difference
will dominate the stochastic part (\ref{Ltb}). We verified this 
by putting $R$ to zero, leaving everything else unchanged. The result 
is shown as the solid line in Fig. \ref{fig:phi-L}, which is a perfect
fit to the fluctuating data of the stochastic cascade.

Our first approximation will thus be to include only the deterministic
part of the Langevin model in our analytical description. 
A second approximation is of a more technical nature. It is seen in 
Fig. \ref{fig:phi-L} that the sequence of level energies form a reasonable 
approximation of a continuous curve. This motivates us to pass to 
a continuum limit, where the ratio of length scales between two levels 
approaches one, leading to a less cumbersome description in terms of 
differential equations.
We introduce $\lambda$ as the ratio of length scales, which means that 
the length scale on level $\ell$ is $r = \lambda^{-\ell} L$.
As $\lambda$ approaches 1, $E_{\ell}$ can be written as a continuous 
variable $E = E(r)$. Replacing $2^{-\ell} L$ by $\lambda^{-\ell} L$
in (\ref{enbal}) and (\ref{Lt}), and 
performing the limit $\lambda \rightarrow 1$, 
one ends up with 
\begin{equation}
  \label{cont}
\partial_t E(r,t) = \bar{D} r \partial^2_r E^{3/2}(r,t) ,
\end{equation}
where $\bar{D} = D (\ln\lambda)^2$ is a rescaled coupling constant.
The equation of motion (\ref{cont}) is the one our subsequent analytical 
description is based on.

We now look for self-similar solutions of (\ref{cont}) of the 
form
\begin{equation}
  \label{cscal}
  \tilde E({\tilde r},\tilde\tau)={\tilde r}^{\alpha}\Phi(\xi)\quad ,
   \xi = {\tilde r}/{\tilde \tau}^{\beta} \quad,
\end{equation}
in direct analogy to (\ref{scalform}). Plugging this into the 
equations of motion, we find that the explicit dependence on
$\tilde r$ is eliminated by demanding that $\beta = 2/(2-\alpha)$,
so we recover the scaling relation (\ref{scaling}). As a result,
we are left with a similarity equation which depends on 
$\xi = \tilde r/{\tilde \tau}^{\beta}$ alone:
\begin{eqnarray}
  \frac{2}{2-\alpha}\xi^{2-\alpha/2}\Phi'(\xi) &=& \frac{3}{2}\bar{D}
  \bigg[\alpha(\frac{3}{2}\alpha-1)\Phi^{3/2}(\xi)
    +3\alpha\xi\Phi^{1/2}(\xi)\Phi'(\xi)
    +\frac{1}{2}\xi^{2}\Phi^{-1/2}(\xi)\Phi'^{2}(\xi) \nonumber\\ 
    & &\hspace{2cm}+\xi^{2}\Phi^{1/2}(\xi)\Phi''(\xi)\bigg]\quad . 
    \label{simphi}
\end{eqnarray}
In the following section we will show that (\ref{simphi}) possesses
a unique physical solution, which fixes both the similarity 
function $\Phi$ and the exponent $\alpha$.

%%%%%%%%%%%%%%%%%
% section 5     %
%%%%%%%%%%%%%%%%%

\section{Selection}
\label{sec:selec}

Since (\ref{simphi}) is of second order, one needs two 
initial conditions to uniquely fix the solution. 
As noted earlier, $\Phi$ asymptotes to a constant at infinity.
Since scales have been normalized according to (\ref{refscal}),
this constant is one, leaving us with the boundary conditions
\begin{equation}
           \left.\begin{array}{lcl}
            \Phi(\xi) &\rightarrow & 1\\
            \Phi'(\xi) &\rightarrow& 0\\
            \end{array}\right\} \xi \rightarrow \infty \quad.
\label{boundary}
\end{equation}
The coupling strength $\bar{D}$ can be eliminated by the transformation
\begin{equation}
  \label{trans}
\zeta = \bar{D}^{2/(\alpha-2)} \xi \quad.
\end{equation}
This means that for large coupling strengths the position of the front
moves towards larger $\xi$. Thus a given length scale 
is reached earlier, as to be expected on physical grounds. 
For our discussion of universal solutions we will consider the
equation in the independent variable $\zeta$, where $\bar{D}$ 
has been eliminated.

\begin{figure}[H]
  \begin{center}
    \leavevmode
    \epsfsize=0.5 \textwidth
    \epsffile{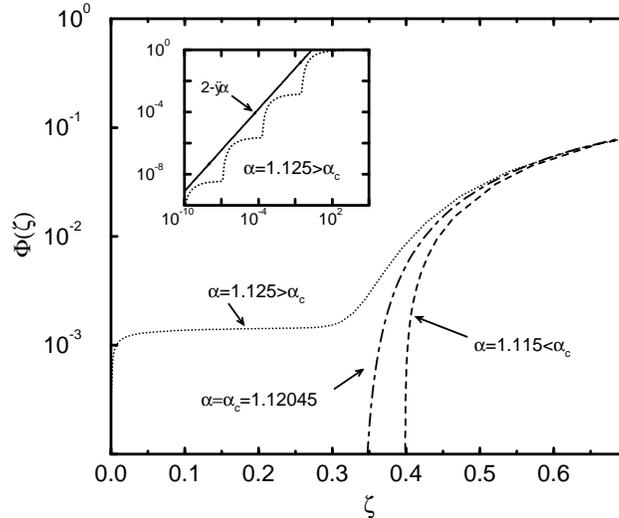}        
    \caption{Solutions of the similarity equation (\protect{\ref{simphi}})
      for three different $\alpha$. For $\alpha$ greater than a 
      critical value $\alpha_c = 1.12045$, the tip of the front
      becomes fractal. The inset shows a blow-up of this behavior
      on a logarithmic scale.
             }
    \label{fig:sel}
  \end{center}
\end{figure}

To find explicit solutions, we expand the similarity equation around
$\zeta = \infty$. Using the boundary conditions, this leads to an
asymptotic expansion
\begin{equation}
  \label{inf}
\Phi(\zeta) = \sum_{i=0}^{\infty} b_i \zeta^{-i/\beta} \quad,\quad
b_0 = 1 \quad,
\end{equation}
where the coefficients $b_i$ depend on $\alpha$ alone. Given a
sufficiently large $\zeta_{init}$, (\ref{inf}) can be used to generate
an initial condition at $\zeta_{init}$. This initial condition then
allows to integrate the similarity equation numerically 
towards small $\zeta$. Hence there is a unique 
solution {\it for each} $\alpha$, while we expect the partial 
differential equation (\ref{cont}) to select a unique $\alpha$.
To understand this, we now look at the behavior of 
solutions for different $\alpha$. Three cases arise, which 
are shown in Fig. \ref{fig:sel}.
If $\alpha$ is smaller than a critical value $\alpha_c = 1.12045\pm 10^{-5}$
the profile ends in a sharp front, similar to the front observed 
in simulations both of the REWA and the Langevin cascade.
The profile is zero below a front position $\zeta_0$, for $\zeta > \zeta_0$
$\Phi$ behaves asymptotically like $\Phi \sim (\zeta - \zeta_0)^{2/3}$.
This corresponds to a local expansion of the form
\begin{equation}
   \label{exp1}
  \Phi(\zeta) = \sum_{i=0}^{\infty} a_i (\zeta - \zeta_0)^{(2+i)/3} \quad.
\end{equation}

If on the other hand $\alpha$ is larger than $\alpha_c$, the front becomes 
unstable and levels off to form a flat plateau. As shown in more
detail in the appendix (see also the inset in Fig. \ref{fig:sel}),
this plateau dips down again at a smaller 
value of $\zeta$, only to form another plateau as the second front
becomes unstable. This process repeats itself, to form a fractal 
tip which asymptotes to zero. Clearly this is not an acceptable 
solution, at the very least because it would correspond to energy
being transported instantly across all levels. 

This indicates that there is something special about solutions 
at $\alpha = \alpha_c$, which separates the region of sharp and
fractal fronts. Indeed, a more careful analysis, which is detailed
in the appendix, reveals that at the critical $\alpha$ the asymptotics
at the front position is now
 \begin{equation}
   \label{exp2}
  \Phi(\zeta) = \sum_{i=0}^{\infty} 
      \bar{a}_i (\zeta - \zeta_0)^{2+i} \quad.
\end{equation}
But although this amounts only to a slight difference in the appearance
of the fronts in Fig. \ref{fig:sel}, the solution at $\alpha_c$ 
is the one which is selected. This comes from an argument similar 
to that advanced for front propagation into an unstable medium
\cite{s88}. Indeed, on a logarithmic scale, i.e. by level number,
the self-similar solution (\ref{cscal}) corresponds to a front 
propagating at a constant speed $\alpha$. In our problem, the 
situation is actually reverse to that of \cite{s88}: of all possible 
solutions, the one with the highest speed will eventually take 
over, while the slow solutions are left behind. On the time 
scale set by the front, they no longer move, and thus drop out of the 
problem. This explains the universality observed earlier: 
independent of initial conditions, only one front with a given exponent 
$\alpha$ is observed. We also checked the validity 
of our selection argument directly, by simulating the Langevin 
cascade (\ref{enbal}), (\ref{Lt}) with $R = 0$ 
for smaller and smaller scaling factors $\lambda$. Extrapolating
to $\lambda = 1$, we were able to confirm the value of 
$\alpha_c$ to five decimal places.

In Table \ref{tab:alpha} we summarize some of the values of the 
exponent $\alpha$ obtained for different cascades. The REWA cascade
with $26$ modes, which we considered throughout this paper, is called 
``small cascade'' here, to distinguish it from another mode selection
with $74$ modes. The Langevin cascade with the same scale factor 
$\lambda = 2$ as the REWA cascades gives a somewhat larger value. 
\begin{table}[htbp]
  \begin{center}
    \leavevmode
    \begin{tabular}[t]{lcc}
      & $\alpha$
      \\ \hline
       small cascade, N=26 & $1.12 \pm 0.01$ \\
       large cascade, N=74 & $1.17 \pm 0.01$  \\
       Langevin cascade, $\lambda = 2$ & $1.24 \pm 0.01$ \\
       similarity equation & $1.12045 \pm 10^{-5}$
     \end{tabular}
  \end{center}
  \caption{Compilation of different values of the exponent 
     of the shell energies $\alpha$. The first two values 
     are from simulations of the REWA cascade with two different
     wave vector sets. The third line refers to the model 
     equations (\protect{\ref{enbal}}),(\protect{\ref{Lt}}), 
     the last line is the result of our similarity theory.
              }
  \label{tab:alpha}
\end{table}
\noindent
However, the overall variation of the exponent $\alpha$ 
is only in the order of 10\%. This strongly supports the 
claim that our analytical theory has captured the physical mechanism behind 
the selection of a singular front for all the models considered. 

\begin{figure}[H]
  \begin{center}
    \leavevmode
    \epsfsize=0.5 \textwidth
    \epsffile{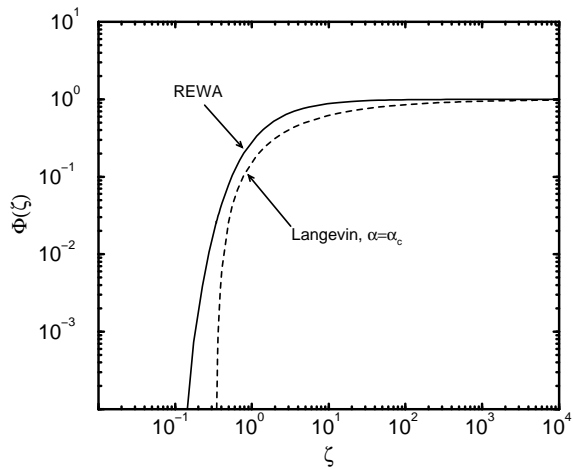}        
    \caption{Comparison between the similarity theory and 
            and the numerical simulation of the REWA cascade.
            The free parameter $\bar{D}$ of the similarity 
            equation has been determined from an independent 
            measurement of the turbulent state.
             }
    \label{fig:comp}
  \end{center}
\end{figure}

Another important quantity is the position of the front, which 
for the similarity solution is found to be 
\begin{equation}
  \label{rc}
  {\tilde r}_c = \bar{D}^{2/(2-\alpha)} \zeta_0 {\tilde \tau}^{\beta} .
\end{equation}
Once again we see that for large $\bar{D}$ a given length 
is excited at earlier times. The coupling $\bar{D}$ is the 
only parameter to be determined for a comparison between 
theory and simulation of the REWA cascade.
We adjusted $\bar{D}$ such that the average 
energy in a turbulent state agrees with the value determined for
the REWA cascade \cite{ue96}. Thus we are able to predict $\Phi$ without
adjustable parameters. The comparison between the solution of the 
similarity equation at $\alpha = \alpha_c$ and the numerical 
simulation of the REWA cascade is shown in Fig. \ref{fig:comp}.
\noindent
The agreement is quite good, although there is some difference 
in the shape of the front. It makes sense that for this shape,
which describes the excitation of Fourier modes, the detailed  
coupling structure of modes matters to some degree. This shows that the 
formation of a singularity in a very complicated system of 
coupled Fourier modes is indeed described by a very simple physical 
principle: the tendency of the cascade to establish equipartition 
of energy.

%%%%%%%%%%%%%%%%%
% conclusion    %
%%%%%%%%%%%%%%%%%

\section{Discussion}
\label{sec:d}

Roughly speaking, two different methods have been used to 
numerically treat the formation of singularities in the 
Euler equation. One relies on a Fourier representation of
the velocity field \cite{b83,b92}, the other tries to track
the vorticity in real space \cite{s85,gs91}. Given a simulation 
with perfect resolution, the results should be the same, but 
in effect there are two different physical pictures underlying 
the two approaches. A spectral approach is preferable if a 
singularity generically results from the interaction of
Fourier modes, as suggested originally by Onsager \cite{o49}.
A spatial picture would show a uniformly (multi)-fractal structure. 

Vorticity dynamics, on the other hand, would be more useful if a
singularity results from a specific spatial structure, like the
meeting of two antiparallel vortex tubes \cite{s85}. Other, more
complicated structures have been proposed in two-dimensional,
axisymmetric flow \cite{gs91,ps92}.

Since the spatial resolution of the REWA cascades studied here is 
quite low, it is hard to speculate about the physics of vortex 
tubes. We must rely on the spectral picture being relevant,
which was analyzed in detail by Brachet et al. \cite{b83}.
They find that the large 
wavenumber part of the energy spectrum is well described by 
\begin{equation}
  \label{brachet}
E(k,t) = a k^{-n(t)} e^{-2\delta(t)k} \quad.
\end{equation}
This is very similar to the form of the energy distribution observed
for our case. However, an important difference is that (\ref{brachet}) 
predicts excitation of all scales, while in the REWA model the
energy drops to zero. This difference, which becomes noticeable 
for energies smaller than those shown in Fig. \ref{fig:phi}, comes
from the absence of non-local interactions in the REWA model, which would
excite small scales instantly.
Since the energy $E_{\ell}$ of a shell represents the 
energy spectrum integrated over an octave in wavenumber, $n(t)$ is
to be identified with the exponent $\alpha + 1$. If in addition 
$\delta(t)$ was chosen to behave like $\tau^{\beta}$, one would 
end up with our scaling form (\ref{cscal}). 

However, although $n(t)$ converges to a value close to
$4$ \cite{b83}, this is much larger than the value $\alpha+1 = 2.12$
we find. Other workers \cite{k93} find $n$ close to $3$ 
asymptotically. This is closer to our value, but still at the 
boundary where the largest scales would move fastest, according
to our estimate $\tau_r = r^{1 - \alpha/2}$. Our simulations are
self-consistent since for $\alpha \ge 2$ {\it nonlocal} interactions
would become dominant \cite{o49,ey94,ey95}, which have not been included
in our description. It would of course be of great importance to 
investigate whether local or non-local interactions are dominant 
for the full Euler equation.

A second significant 
difference between the results of \cite{b83} and our work is that 
$\delta(t)$ does not go to zero in finite time, but rather behaves 
like 
\begin{equation}
   \label{delta}
\delta(t) = \delta_0 e^{-t/T} \quad.
\end{equation}
This corresponds to a singularity only in infinite time. All
results of \cite{b83} were later confirmed in \cite{b92} using
greater resolution and more general initial conditions.
But of course there is also a good possibility that if greater 
resolution allows to continue the simulation still further, 
a crossover from (\ref{delta}) to a finite time singularity 
is seen. 

A definite answer whether singularities of the three-dimensional 
Euler equation exist can only be given by considering 
the full equations. Nevertheless, our study shows that structures exist
in inviscid flow which show blow-up in finite time. They are consistent
with the divergence of the vorticity like $\tau^{-1}$, and the 
existence of a local cascade. The scaling structure 
of the singularity found here can thus serve as a guideline for 
further studies of the full equations. The more general structures 
possible in the fully resolved flow can of course be more 
singular than ours, so we expect the structures appearing in our 
mode-reduced systems to be dominated by even more singular contributions. 
Conversely, there is also the possibility that the modes not taken 
into account in our study will interact with the reduced mode system 
to keep it from becoming singular. One possible way to study this 
would be to selectively take  non-local interactions into account,
to find out what modifications they imply for our analysis. 

A particularly intriguing aspect of the present work is the novel 
way the scaling exponent is selected as the ``speed'' of a marginally
stable solution. This adds another variety to the existing mechanisms
which determine the scaling exponents of singularities which are not
determined from dimensional arguments. Other selection mechanisms 
for this scaling behavior of the second kind (in Barenblatt's
\cite{b79} terminology) are found in \cite{g42} and in \cite{p95}. 

Formally, the selection mechanism is quite similar to that of 
marginally stable solutions of equations of the form \cite{s88}
\begin{equation}
   \label{fk}
\partial_t \phi = \partial^2_x \phi + F(\phi) \quad.
\end{equation}
However, solutions to this equation do not drop to zero at
a finite value of $x$ but rather decay exponentially away from the front. 
For this reason we are not able to repeat the linear stability analysis 
presented for example in \cite{s88}, 
since the front of our solution is 
very steep. In that respect it is more similar to solutions of the 
porous medium equation \cite{a86}, which in one dimension reads 
\begin{equation}
  \label{p}
 \partial_t u = \partial^2_x u^m \quad.
\end{equation}
Solutions to this equation form a front which drops to zero,
like ours. However, although for $m = 3/2$ (\ref{p}) looks 
quite similar to (\ref{cont}), the factor $r$ in (\ref{cont})
represents a very serious complication. This is because its value 
at the front goes to zero at the singularity, giving a very 
singular diffusion constant. Thus again we are not able to 
carry over the mathematically rigorous results known for the porous 
medium equation. Still we believe that the combination of numerical 
evidence and analysis of the similarity equation conclusively 
demonstrates that the marginal solution is indeed the one selected 
for our equation (\ref{cont}). This adds to the generality of the 
marginal stability concept. 

Finally, we would like to stress the fact that the cascades studied 
in this paper give a coherent picture of {\it both} Euler singularities 
and fully developed turbulence. In fact the same is true for a 
scalar shell model recently studied by Dombre and Gilson \cite{dg95}. 
They also found unique singular solutions of the inviscid 
equations, but whose spectrum is {\it less} steep than Kolmogorov's.
In addition, they propose a connection between inviscid 
singularities and intermittent fluctuations. 
We started to explore this connection
in the second section, but this has
to be pursued further. We suspect the situation will be quite different 
from that of the scalar shell model, because our singular spectrum is 
steeper than that of a stationary cascade. 
In particular, it would be 
interesting to study the interaction between the most singular solutions 
and viscosity, leading to a turbulent state. Since the most singular
solution with $\alpha = \alpha_c$ is regularized by viscosity, 
other solutions with $\alpha < \alpha_c$ will become relevant 
as well. So it may be the whole {\it spectrum} of singular solutions
which is relevant to the turbulent state.

%%%%%%%%%%%%%%%%%%%
% acknowledgments %
%%%%%%%%%%%%%%%%%%%

\acknowledgments
We are grateful to R. Grauer for enlightening discussions and to 
R. Graham for his continued support. J. Krug made helpful comments on the 
manuscript.
This work is supported by the Sonderforschungsbereich 237 (Unordnung und
grosse Fluktuationen).

%%%%%%%%%%%%%%%%%%%%%%%%%%%%%%%%%
% Appendix: analytical solution %
%%%%%%%%%%%%%%%%%%%%%%%%%%%%%%%%%

\appendix
\section*{Similarity solutions}

Here we discuss the transition of solutions of the similarity 
equation (\ref{simphi}) from regular to fractal tips in more detail.
The similarity variable is rescaled according to 
$\zeta=\bar D^{2/\alpha-2}\xi$. 

\begin{figure}[H]
  \begin{center}
    \leavevmode
    \epsfsize=0.5 \textwidth
    \epsffile{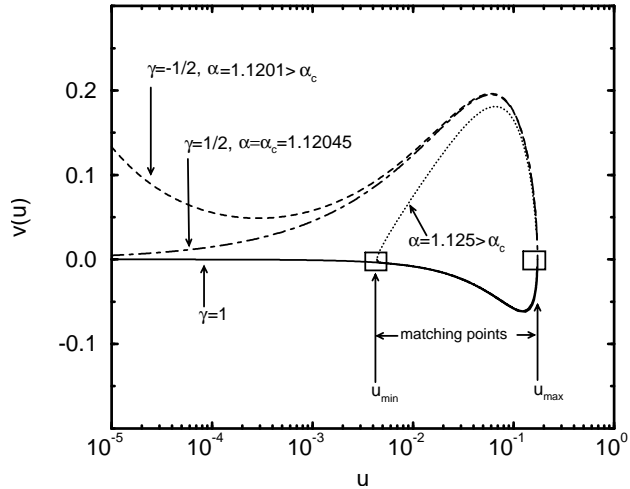}            
  \end{center}
  \caption{Solutions of the first-order version of the similarity
    equation. The three different types of solutions correspond 
    directly to Fig. \protect{\ref{fig:sel}}. The fractal tip, 
    observed for $\alpha > \alpha_c$ corresponds to a limit cycle.
      }
  \label{fig:locsol}
\end{figure}

The invariance of (\ref{simphi}) under scale transformations 
$\zeta\to\mu\zeta$, $\Phi\to\mu^{2-\alpha}\Phi$ can be used to transform 
it to a first order equation
\begin{equation}
  \label{simv}
  v'(u) = \frac{1}{u^{1/2}v(u)}\left[
  \frac{4}{3}u+\frac{4}{6-3\alpha}v(u)-4u^{3/2}-5u^{1/2}v(u)-\frac{1}{2}
    u^{-1/2}v^{2}(u)\right]
\end{equation}
with 
\begin{eqnarray}
  \label{uv}
&& u(\zeta)=\zeta^{\alpha-2}\Phi(\zeta) \nonumber \\
&& v(u)=\zeta u'(\zeta) \quad.
\end{eqnarray}
Primes always refer to derivatives with respect to the argument.
It follows from (\ref{uv}) that $u$ goes to zero both for 
$\zeta$ going to infinity and $\Phi$ going to zero. Thus all 
solutions start out at $u=0$, corresponding to $\zeta=\infty$,
shown by the full line in Fig. \ref{fig:locsol}. To form a regular tip,
they must return to $u = 0$ (dashed and dot-dashed lines), which 
is the case for $\alpha \le \alpha_c$. For $\alpha > \alpha_c$,
on the other hand, $v(u)$ enters a limit cycle, which corresponds to the 
fractal tip (dotted line).

Equation (\ref{simv}) becomes singular for $u = 0$ or $v=0$, thus
only local solutions exist either in the domain 
$0<u<u_{max}$ or $u_{min}<u<u_{max}$ with $v(u_{max/min})=0$.
To form a complete solution, local solutions have to be matched at 
the points indicated in Fig. \ref{fig:locsol}.

Three different expansions exist around the point $u = 0$:
\begin{eqnarray}
  \label{v}
&&  v(u)=u^{\gamma}\sum_{i=0}^{\infty} c_{i}^{(\gamma)}u^{i/2} \nonumber \\
&&  \gamma = 1, -\frac{1}{2}, \frac{1}{2} \quad,
\end{eqnarray}
which corresponds directly to the three expansions 
(\ref{inf}), (\ref{exp1}), and (\ref{exp2}) of $\Phi$.
The coefficients $c_i^{(1)}$, which describe the behavior of 
$\Phi$ at infinity, are determined recursively from
\begin{eqnarray}
  \label{c1}
&&  c_0^{(1)} = \alpha - 2 \quad, \nonumber \\
&&  c_{i}^{(1)} =
  \frac{3(2-\alpha)}{4} \left (4\delta_{i1}+5c_{i-1}^{(1)}+\sum_{j=0}^{i-1}
  \left (\frac{3}{2}+\frac{j}{2}\right ) 
      c_{j}^{(1)}c_{i-j-1}^{(1)}\right )\quad .
\end{eqnarray}
The case $\gamma = 1/2$, which corresponds to a tip of the form
$\Phi(\zeta)=\zeta_{0}^{-\alpha}(c_{0}^{(1/2)})^{2}(\zeta-\zeta_{0})^{2}/4$
is determined by  
\begin{eqnarray}
  \label{c2}
&&  c_0^{(1/2)} = \frac{4}{3(2-\alpha)} \nonumber \\
&&  c_{i}^{(1/2)} = \frac{-2\delta_{i1}+6\delta_{i2}+15c_{i-1}^{(1/2)}/2+
    \sum_{j=1}^{i-1}\left(3/2+3j/4\right)c_{j}^{(1/2)}c_{i-j}^{(1/2)}}
  {2/(2-\alpha)-3\left(1+i/4\right)c_{0}^{(1/2)}}\quad .
\end{eqnarray}
This is the critical case. On the other hand, solutions with 
$\alpha < \alpha_c$ have a tip 
$\Phi(\zeta) =
\zeta_{0}^{1-\alpha}(3c_{0}^{(-1/2)}/2)^{(2/3)}(\zeta-\zeta_{0})^{2/3}$.
In that case the constant $c_0^{(-1/2)}$ is open and the higher order 
coefficients are calculated from
\begin{equation}
  \label{c3}
  c_{i}^{(-1/2)} = \frac{4}{3ic_{0}^{(-1/2)}}
  \left(2\delta_{i5}-6\delta_{i6}+\frac{2}{2-\alpha}c_{i-2}^{(-1/2)}-
       15c_{i-3}^{(-1/2)}/2-
    \sum_{j=2}^{i-2} (3j/4)c_{j}^{(-1/2)}c_{i-j}^{(-1/2)}\right)\quad .
\end{equation}
All expansions (\ref{c1})-(\ref{c3}) are asymptotic in nature, but are  
extremely good everywhere except in a small neighborhood of $u_{max}$.
To do the matching at $u_{max}$, equation (\ref{simv}) has to be 
integrated numerically.

The local behavior at $u_{max}$ is
\begin{equation}
  \label{vum}
  v(u) = C \left(u_{max}-u\right)^{1/2} \mbox{ with } C=\pm
  \left(8(u_{max}-\frac{1}{3}u_{max}^{1/2})\right)^{1/2} \quad,
\end{equation}
where the solid line of Fig. \ref{fig:locsol} corresponds to 
negative $C$, the other branches to positive $C$. At a given $\alpha$, 
for each of the three branches $\gamma = 1, -1/2$, and $1/2$ one 
can extrapolate to the asymptotic behavior (\ref{vum}) to 
determine $u_{max}$. A matching of the branch $\gamma = 1$ on the 
branch $\gamma = 1/2$ is only possible for one $\alpha$, which 
is the critical $\alpha = \alpha_c = 1.12045$. For $\gamma = -1/2$
the free parameter $c_0^{(-1/2)}$ can be used to match $u_{max}$,
and this turns out to be possible only for $\alpha < \alpha_c$.

If $\alpha > \alpha_c$, there is no branch which returns to $u = 0$,
and the solution has to be continued numerically through the matching point.
The resulting branch goes to $v = 0$ at a finite $u = u_{min}$, 
where it can be matched onto the original solution with $\gamma =1$.
This means the solution ends up on a limit cycle. From the 
definition of $u$ (cf. (\ref{uv})) it is clear that the sequence
of tips of $\Phi(\zeta)$ approximates a power law 
$\Phi \sim \zeta^{2 - \alpha}$, as shown in the inset of 
Fig. \ref{fig:sel}.

%%%%%%%%%%%%%%%%%
% bibliography  %
%%%%%%%%%%%%%%%%%


\begin{references}

\bibitem{o49} L.~Onsager, ``Statistical hydrodynamics'',
  {Nuovo Cimento} {\bf 6}, 279 (1949)

\bibitem{b83} M.E. Brachet et al., ``Small-scale structure of the 
Taylor-Green vortex'',
  {J. Fluid Mech.} {\bf 130}, 411 (1983)

\bibitem{s85} E. D. Siggia, ``Collapse and amplification
of a vortex filament'',
Phys. Fluids {\bf 28}, 794 (1985)

\bibitem{ps90} A. Pumir and E. D. Siggia, ``Collapsing
solutions of the 3-D Euler equations'',
Phys. Fluids A {\bf 2}, 220 (1990)

\bibitem{gs91} R. Grauer and T. C. Sideris, ``Numerical Computation
of 3D Incompressible Ideal Fluids with Swirl'',
Phys. Rev. Lett. {\bf 67}, 3511 (1991)

\bibitem{ps92} A. Pumir and E. D. Siggia, ``Development of
singular solutions of the axisymmetric Euler equation'',
Phys. Fluids A {\bf 4}, 1472 (1992)

\bibitem{smo93} M. J. Shelley, D. I. Meiron, and
S. A. Orszag, ``Dynamical aspects of vortex reconnection of 
perturbed anti-parallel vortex tubes'',
J. Fluid Mech. {\bf 246}, 613 (1993)

\bibitem{k93} R.~M.~Kerr, ``Evidence for a singularity of the 
three-dimensional, incompressible Euler equations'', 
Phys. Fluids A {\bf 5}, 1725 (1993)

\bibitem{b92} M.E. Brachet et al., ``Numerical evidence of smooth
  self-similar dynamics and possibility of subsequent collapse for 
  three-dimensional ideal flows'',
  {Phys. Fluids A} {\bf 4}, 2845 (1992)

\bibitem{e93} J.~Eggers, ``Universal pinching of 3D
Axisymmetric Free-Surface Flow'',
Phys. Rev. Lett. {\bf 71}, 3458 (1993)

\bibitem{k41} A.~N.~Kolmogorov, ``The local structure of turbulence in
  incompressible viscous liquids'',
  {Dokl. Akad. Nauk SSSR} {\bf 30}, 9 (1941)

\bibitem{eg91} J.~Eggers and S.~Grossmann, ``Does deterministic chaos imply
  intermittency in fully developed turbulence?'',
  {Phys. Fluids A} {\bf 3}, 1985 (1991)

\bibitem{gl94} S.~Grossmann and D.~Lohse, ``Scale resolved intermittency in
  turbulence'', 
Phys. Fluids {\bf 6}, 611 (1994)

\bibitem{ue96} C.~Uhlig and J.~Eggers, ``Local coupling of 
shell models leads to anomalous scaling'',
Z. Phys. B {\it to be published} (1996)

\bibitem{nelkin89} M.~Nelkin, ``What do we know about self-similarity 
in fluid turbulence?'', J. Stat. Phys. {\bf 54}, 1 (1989).

\bibitem{e94} J.~Eggers, ``Multifractal scaling from nonlinear turbulence
  dynamics: analytical methods'', 
  {Phys. Rev. E} {\bf 50}, 285 (1994)

\bibitem{obukhov71} A.~M.~Obukhov, ``Some general properties of 
equations describing the dynamics of the atmosphere'', Atmos. Ocean. Phys.
 {\bf 7}, 471 (1971).

\bibitem{gledzer73} E.~B.~Gledzer, ``Systems of hydrodynamic type
admitting two quadratic integrals of motion'', Sov. Phys. Dokl.
 {\bf 18}, 216 (1973).

\bibitem{yamada87} M.~Yamada and K.~Ohkitami, ``Lyapunov spectrum
of a chaotic model of three-dimensional turbulence'', J. Phys. Soc. Jpn
  {\bf 56}, 4210 (1987).

\bibitem{jensen91} M.~H.~Jensen, G.~Paladin, and A.~Vulpiani, 
``Intermittency in a cascade model for three-dimensional turbulence'',
Phys. Rev. A  {\bf 43}, 798 (1991).

\bibitem{f95} U.~Frisch, ``Turbulence'', 
  (Cambridge University Press, 1st ed., 1995)

\bibitem{bkm84} J.T. Beale, T. Kato, and A. Majda, ``Remarks on 
the breakdown of smooth solutions for the 3D Euler equations'', 
  {Comm. Math. Phys.} {\bf 94}, 61 (1984)

\bibitem{ey94} G. L. Eyink, ``Energy dissipation without 
viscosity in ideal hydrodynamics, I'',
Physica D {\bf 78}, 222 (1994)

\bibitem{ey95} G. L. Eyink, ``Local energy flux and the refined 
similarity hypothesis'',
J. Stat. Phys. {\bf 78}, 335 (1995)

\bibitem{b79} Barenblatt, G. I., {\it Scaling, self-similarity, and 
intermediate asymptotics}, (Cambridge University Press, 1996)

\bibitem{g42} A. Guderley, ``Starke kugelige und zylindrische 
Verdichtungsst\"osse in der N\"ahe des Kugelmittelpunktes bzw.
der Zylinderachse'', Luftfahrtforschung {\bf 19}, 302 (1942)

\bibitem{p95} D. T. Papageorgiou, ``On the breakup of viscous 
liquid threads'', Phys. Fluids {\bf 7}, 1529 (1995)

\bibitem{s88} W. van Saarloos,
  ``Front propagation into unstable states: Marginal stability 
as a dynamical mechanism for velocity selection'',
{Phys. Rev. A} {\bf 37}, 211 (1988)

\bibitem{a86} D. G. Aronson, ``The porous medium equation'',
  in : {\em Nonlinear Diffusion Problems},
 edited by A. Fasano and M.Primicario, Lecture Notes in Mathematics,
  (Springer, 1986)

\bibitem{dg95} T. Dombre and J.-L. Gilson, ``Intermittency, 
chaos and singular fluctuations in the mixed Obukhov-Novikov 
shell model of turbulence'',
preprint, October 1995

\end{references}
\end{document}